\documentclass[twoside,twocolumn,english,pra,twocolumn,aps,superscriptaddress,showpacs]{revtex4-1}
\usepackage[T1]{fontenc}
\usepackage[latin9]{inputenc}
\usepackage{color}
\usepackage{babel}
\usepackage{textcomp}
\usepackage{amsmath}
\usepackage{appendix}
\usepackage{amsthm}
\usepackage{graphicx}
\usepackage{booktabs}
\usepackage{longtable}
\usepackage{multirow}
\usepackage[unicode=true,pdfusetitle,
bookmarks=true,bookmarksnumbered=false,bookmarksopen=false,
breaklinks=false,pdfborder={0 0 0},backref=false,colorlinks=true,citecolor=blue]
{hyperref}
\usepackage{breakurl}
\usepackage{epstopdf}
\usepackage{bm}

\makeatletter
\theoremstyle{plain}

\makeatother

\preprint{This line only printed with preprint option}
\begin{document}

\title{Dynamically Characterizing the Structures of Dirac Points via Wave Packets}

\author{Dan-Dan Liang}
\affiliation {Key Laboratory of Atomic and Subatomic Structure and Quantum Control (Ministry of Education), School of Physics, South China Normal University, Guangzhou 510006, China}

\affiliation {Guangdong Provincial Key Laboratory of Quantum Engineering and Quantum Materials, South China Normal University, Guangzhou 510006, China}

\affiliation {Guangdong-Hong Kong Joint Laboratory of Quantum Matter, Frontier Research Institute for Physics, South China Normal University, Guangzhou 510006, China}

\author{Xin Shen}
\email{shenx@cjlu.edu.cn}
\affiliation{College of Sciences, China Jiliang University, Hangzhou 310018, China}

\author{Zhi Li}
\email{lizphys@m.scnu.edu.cn}
\affiliation {Key Laboratory of Atomic and Subatomic Structure and Quantum Control (Ministry of Education), School of Physics, South China Normal University, Guangzhou 510006, China}

\affiliation {Guangdong Provincial Key Laboratory of Quantum Engineering and Quantum Materials, South China Normal University, Guangzhou 510006, China}

\affiliation {Guangdong-Hong Kong Joint Laboratory of Quantum Matter, Frontier Research Institute for Physics, South China Normal University, Guangzhou 510006, China}

\date{\today}

\begin{abstract}

Topological non-trivial band structures are the core problem in the field of topological materials. In this paper, we investigate the topological band structure in a system with controllable Dirac points from the perspective of wave packet dynamics. By adding a third-nearest-neighboring coupling to the graphene model, additional pairs of Dirac points emerge. The emergence and annihilation of Dirac points result in hybrid and parabolic points, and we show that these band structures can be revealed by the dynamical behaviors of wave packets. Particularly, for the gapped hybrid point, the motion of the wave packet shows a one-dimensional \emph{Zitterbewegung} motion. Furthermore, we also show that the winding number associated with the Dirac point and parabolic point can be determined via the center-of-mass and spin texture of wave packets, respectively. The results of this work could motivate new experimental methods to characterize the system's topological signatures through wave packet dynamics, which may also find application in systems of other exotic topological materials.  

\end{abstract} 

\maketitle

\emph{Introduction---}Ever since the birth of quantum mechanics, the solid band theory provides the fundamental explanation for metals, insulators, and semiconductors~\cite{Ashcroft1976}. Nevertheless, both the theoretical and experimental progress in condensed matter physics have greatly extended the classification of materials and deepened the concept of phase transition in recent years~\cite{Wen2017,topo_material}. Among those novel physics are the topological insulators and superconductors~\cite{hasan_ti,xlqi_ti} or semimetals~\cite{Diracmetal1,Diracmetal2,weylfermion}, which trigger tremendous investigations in quantum transport~\cite{Lu2017}, synthetic topological quantum matter~\cite{Yang2015,Peri2020,topo_photonics,topo_coldatom2,Monroe2021}, and hence potential application in quantum computation~\cite{topo_computation}.  Among topological materials, the band structure of a Dirac point is ubiquitous in indicating the topological phase transition or band inversion~\cite{Shen2018,Shen2022}.  The band's topology can be characterized by an invariant~\cite{Chiu2016}. Much effort has been paid to the exploration of topological states via measurements of topological invariants or edge states, especially in the experimental simulation of topological matters in various platforms such as optical crystals~\cite{topo_photonics}, trapped ions~\cite{Monroe2021}, and ultracold atoms~\cite{topo_coldatom2}.  

In this paper, we report the investigation of band structure via wave packet dynamics that uncovers the topological property encoded in the band. The corresponding Dirac points can be simulated in an optical lattice in the form of low-energy excitation~\cite{Tarruell2012}. The number of Dirac points is restricted to even numbers with opposite winding numbers or chirality~\cite{Duca2015}due to the Fermi doubling theorem~\cite{fermi_doubling}.  By tuning the intensity of lasers, the motion of Dirac points can be induced with the emergence or annihilation of Dirac points. We focus on a prototypical graphene model hosting the Dirac quasiparticles with the third-nearest-neighboring (N3) coupling considered~\cite{Bena2011,Montambaux2012}. 
The N3 coupling shifts the position of Dirac points and modifies the band structure into a topologically trivial hybrid point or a topologically non-trivial parabolic point with a higher winding number~\cite{Montambaux2018}. A wave packet can be viewed as a state with a definite (quasi)momentum, therefore its expansion reveals local band structure such as the group velocity and band gap. 
The merging of Dirac points with opposite winding numbers gives rise to a gapped hybrid point, leading to a directed \emph{Zitterbewegung}  (ZB) motion~\cite{schrozb}. The merging process also happens for Dirac points of the same winding numbers, forming a gapless parabolic point of a higher winding number. By a single wave packet's dynamics, we show that the winding number can be inferred from the motion of wave packets' center of mass and spin texture, providing a dynamical framework to characterize the topological band structures~\cite{Cooper2019,Zhang2013,Deng2017,Guo2021}.

\begin{figure} \centering
\includegraphics[width=7.5cm]{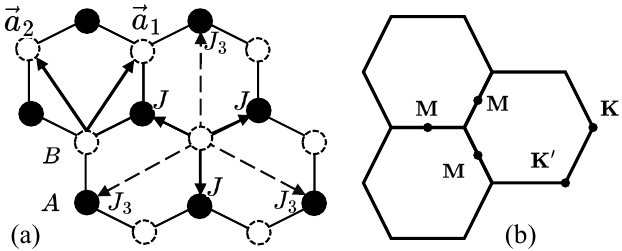}
\caption{(a) The tight-binding model schematic diagram of honeycomb lattice with the nearest and third-nearest neighboring hopping. (b) Brillouin zone and the high symmetry points. }
\label{model}
\end{figure}

\emph{Model and band structures---}Within the tight-binding approximation, the graphene's Hamiltonian can be written as $H=\sum_{mn}c^\dagger_m J_{mn} c_n$, where $c,c^\dagger$ are the creation and annihilation operators of each lattice site, and $J_{mn}$ characterize the hopping between lattices sites. Considering only the nearest neighboring hopping,  gapless Dirac points emerge in the graphene merely at the $\mathbf K$ and $\mathbf K'$ valley. By adding a third-nearest-neighboring coupling to the tight-binding model of graphene as shown in Fig.~\ref{model}, the preserved chiral symmetry allows for the emergence of additional pairs of Dirac points. The corresponding Hamiltonian reads
\begin{equation}
H=J\sum_{i,j=NN}c_{A,i}^\dagger c_{B,j}+J_3\sum_{i,j=N3}c_{A,i}^\dagger c_{B,j}+\text{h.c.},
\end{equation}
where $\rm{NN}$ stands for the hopping between the nearest neighboring sites and $\rm{N3}$ characterizes the hopping between the third nearest neighboring sites with hopping parameters $J$ and $J_3$, respectively. $A$ and $B$ denote two sublattices in the monolayer graphene. By Fourier transformation, the corresponding Bloch Hamiltonian reads
\begin{equation}
\label{eq2}
\mathcal H_{\bm{k}}=
\begin{bmatrix}
0  &  f_{\bm{k}}\\
f_{\bm{k}}^* & 0
\end{bmatrix},
\end{equation}
where $f_{\mathbf{k}}=J(1+e^{i\vec{k}\cdot \vec{a}_1}+e^{i\vec{k}\cdot \vec{a}_2}) + J_3\left[ e^{i\vec{k}\cdot (\vec{a}_1+ \vec{a}_2)}+e^{i\vec{k}\cdot (\vec{a}_1- \vec{a}_2)}+e^{i\vec{k}\cdot (\vec{a}_2- \vec{a}_1)}\right]$. $\vec{a}_1=(\frac{\sqrt{3}a}{2}, \frac{3a}{2})$ and $\vec{a}_2=( -\frac{\sqrt{3}a}{2}, \frac{3a}{2})$ are the elementary vectors of the lattice. $\mathbf k$ is the quasimomentum.  

Due to the $J_3$ term, the energy spectrum is drastically changed. When $J_3$ increases and reaches the critical value $J/3$, a new pair of Dirac points emerge from each of the three inequivalent time-reversal invariant $\mathbf{M}$ points in reciprocal space. 
When further increasing $J_3$, either of the newly emerged pair of Dirac points moves the $\mathbf M$ point to the $\mathbf K$ or $\mathbf K'$ valley and then four Dirac points merge into a single degeneracy. The critical value is $J_3=J/2$, in which case the dispersion becomes quadratic. Near $J_3\approx{J/2}$, the Hamiltonian  in the vicinity of the $\mathbf{K'}$ point can be derived as
\begin{equation}
\mathcal H_K=
\begin{bmatrix}
0 & -\frac{\mathbf{q}^2}{2m^*}+c\mathbf{q}^{\dag}+\Delta\\
 -\frac{\mathbf{q}^{\dag2}}{2m^*}+c\mathbf{q}+\Delta^* & 0
\label{eq3}
\end{bmatrix},
\end{equation} 
where $\mathbf{q} \equiv q_x+iq_y$ and $\mathbf k=\mathbf K'+\mathbf q$. Here, $\mathbf q$ denotes the displacement from $\mathbf K'$ point.  The parameters are $m^*=-4/9J$, $c=3J/2-3J_3$ and $\Delta=0$. When $J_3=J/2$, i.e., $c=0$, the dispersion shows quadratic behavior in both momenta $q_x$ and $q_y$. This differs from the usual parabolic kinetic energy and carries a vorticity. 
\begin{figure*}[tbp] \centering
\includegraphics[width=1\textwidth]{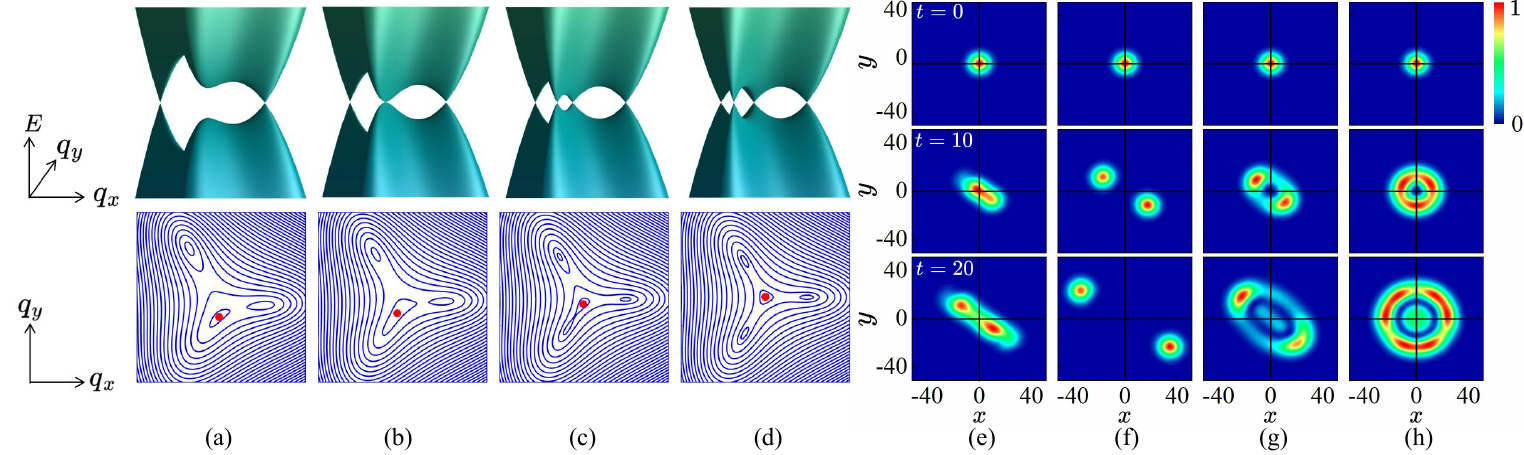}
\caption{ Band structures (left panel) of the Hamiltonian Eq.~\eqref{eq3} and corresponding density profiles $\left|\psi(\mathbf r,t)\right|^2$ (right panel) of the wave packets, with $\Delta=-2\Delta_c,-\Delta_c,-0.5\Delta_c,0$ from (a) to (d) and (e) to (f). The red point marks the center of the wave packet. The units for momentum, position and time are $1/a$, $a$ and $a/c$, respectively. The width of wave packets is $d=5a$. Throughout the simulation $m=1/ac$.  }
\label{fig2}
\end{figure*}
Below, the parameter $\Delta$ is assigned with purely imaginary value $\Delta=ib$ ($b$ is a real number), such that all the band characteristics are included in the Hamiltonian given in Eq.~\eqref{eq3}, i.e., emergence or annihilation of Dirac points with opposite chirality, and also a quadratic dispersion carrying nontrivial vorticity~\cite{Montambaux2018}. The energy spectra are plotted in Fig.~\ref{fig2}. Starting from Fig.~\ref{fig2}(d), when $\Delta = 0$, there are four Dirac cones. Increasing or decreasing $\Delta$ causes two Dirac points to merge, resulting in a linear-quadratic degeneracy and a gapped band, also known as a hybrid point. This can be seen from Fig.~\ref{fig2}(d) to (a).  Below we will discuss the band structures of the model for varying values of $\Delta$, along with their corresponding dynamical behaviors.

\begin{figure}\centering
\includegraphics[width=8.5cm]{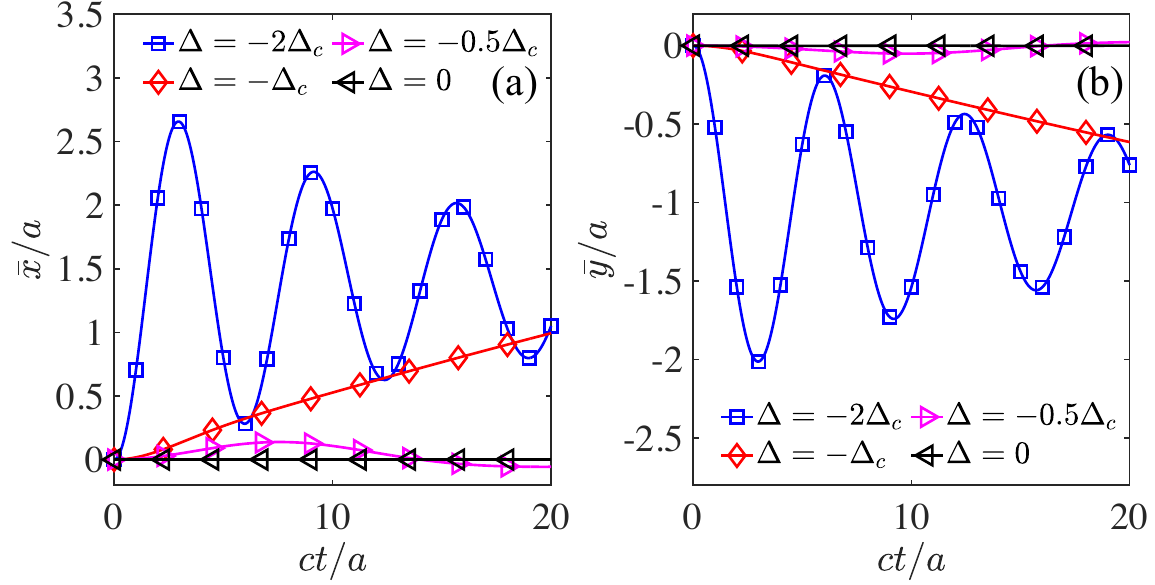}
\caption{(Color online) Analytically (lines) and numerically (symbols) calculated PCM $\bar x$ (a) and $\bar y$ (b) with different values of $\Delta$. The width of the initial wave packet $d=5a$.   }
\label{fig4}
\end{figure}

\emph{The wave packet dynamics---}The wave function for the low-energy quasiparticles satisfies the Schr{\"o}dinger's equation $i \partial_t \psi={\mathcal H}_{K}\psi$ ($\hbar=1$). Throughout the dynamical simulation, we take the initial state as a two-dimensional Gaussian wave packet~\cite{Li2015,Li2016,Shenxin2020}.
The  wave function in momentum space is 
\begin{equation}
\label{eq6}
\mathbf{\psi}_k(t=0)=d\sqrt{\frac{2}{\pi}} e^{-d^2[(k_x-k_1)^2+(k_y-k_2)^2]}\mathbf{\phi}.
\end{equation}
where $d$ is the width of the wave packet, and $k_1(k_2)$ is the initial momentum in the $x(y)$ direction.  $\phi=(c_1,c_2)^T$ is the spinor part, where $T$ stands for the matrix transposition. When the width of a wave packet $d$ is sufficiently large, it is narrowly spread around the point $(k_1,k_2)$ in momentum space, and its time evolution is governed by the Hamiltonian near that point. In order to reveal the local Dirac band structures of Eq.~\eqref{eq3}, $(k_1,k_2)$ take values of the Dirac points marked in Fig.~\ref{fig2}. For a specific demonstration, a monocomponent spinor  $\phi=(1,0)^T$ is considered. To understand the motion behavior,  we proceed to calculate the expectation value of the position operator. By expressing the Hamiltonian Eq.~\eqref{eq3} as $\mathcal H_K=d_x\sigma_x+d_y\sigma_y$,  the position of the center of mass (PCM) is obtained as
\begin{equation}
\label{eq12}
\begin{split}
\bar{x}(t)&=\langle \psi(t)|\hat x|\psi(t)\rangle\\
&=\frac{2d^2}{\pi}\int_{-\infty}^{\infty}\int_{-\infty}^{\infty}\frac{\frac{q_y}{m^*}d_x-(c-\frac{q_x}{m^*})d_y}{2E^2}[\cos(2Et)-1]\\
&\quad\times e^{-2d^2[(k_x-k_1)^2+(k_y-k_2)^2]}dk_xdk_y,\\
\bar{y}(t)&=\langle \psi(t)|\hat y|\psi(t)\rangle\\
&=\frac{2d^2}{\pi}\int_{-\infty}^{\infty}\int_{-\infty}^{\infty}\frac{(c+\frac{q_x}{m^*})d_x-\frac{q_y}{m^*}d_y}{2E^2}[\cos(2Et)-1]\\
&\quad\times e^{-2d^2[(k_x-k_1)^2+(k_y-k_2)^2]}dk_xdk_y,
\end{split}
\end{equation}
where $E=\sqrt{d^2_x+d^2_y}$. The  PCM can also be obtained by directly solving the Schr\"{o}dinger's equation and the wave packet's density profile versus time is shown in Fig.~\ref{fig2}. Both the analytical and numerical results of PCM are plotted in Fig.~\ref{fig4}. From these diagrams, the local band structures can be inferred.  

As shown in Figs.~\ref{fig2}(e), the wave packet exhibits oscillatory motion, corresponding to the gapped band structure after the annihilation of Dirac points. Unlike the 2D massive Dirac quasiparticles, the wave packet here oscillates one-dimensionally due to the linear-quadratic character of the dispersion. When the gap closes, the oscillation ceases and the wave packet splits into two sub-packets moving in opposite directions [see Figs.~\ref{fig2}(b) and (f)]. In this scenario, the one-dimensional motion results from a finite group velocity in the linear direction but zero in the quadratic direction. In Fig.~\ref{fig2}(c), the wave packet shows diffusion due to finite velocities in all directions, corresponding to the band structure of the newly emerged Dirac point. The density distribution is anisotropic because the group velocity in the previous quadratic direction is relatively smaller compared with the velocity in the linear direction. When $\Delta=0$, the four Dirac points are located in momentum space with $C_3$ symmetry, and the initial isotropic wave packet now features the $C_3$ symmetry as well. [see Fig.~\ref{fig2}(d)].

The smaller the width, the more broad the wave packet in momentum space, resulting in an inclusion of more momenta in the dynamics. The dynamics of the local band structure would then be smeared out for a narrow wave packet in position space.   In  Fig.~\ref{fig4} with $\Delta = -2\Delta_c$, the oscillation shows the damping behavior due to the finite width of wave packets. The dynamics can be modeled as a group of oscillators, with each frequency corresponding to an $\mathbf k$-dependent energy gap. To manifest the local band structure's dynamics, a proper width can be inferred from the ZB dynamics with $d\gg c/\Delta\sim a$~\cite{Merkl2008, Zawadzki2011}. Typically in the optical lattice, the trapping frequency is of order $10^2 $Hz, and the lattice constant $\sim \lambda/2$ with $\lambda$ being the laser wavelength. For the trapped $^{87}$Rb Bose condensate~\cite{Meng2023}, for example, the ratio $d/a\sim 17$ is much larger than the numerical simulations in Fig.~\ref{fig2}.  

\begin{figure}[t]\centering
\includegraphics[height=4.2cm]{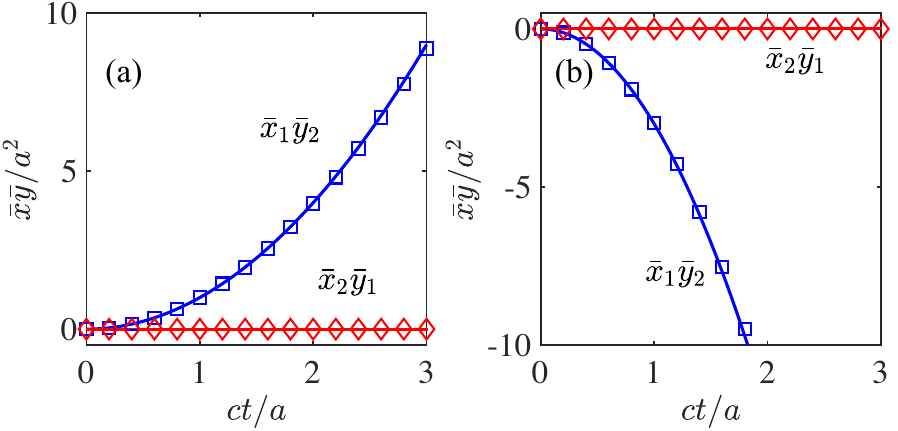}
\caption{(Color online)  Analytically (lines) and numerically (symbols) calculated $\bar x\bar y$ as a function of time $t$.  $(\bar x_1, \bar y_1)$ and $(\bar x_2, \bar y_2)$ are calculated with initial states $|\phi_1 \rangle=1/\sqrt{2}(1,1)^T$ and $|\phi_2 \rangle=1/\sqrt{2}(1,i)^T$, respectively. For (a) and (b), the momentum center of the wave packet is located at the Dirac points  $(0,0),(2m^*c,0)$, respectively. The width of the initial wave packet $d=15a$.   }
\label{fig6}
\end{figure}

\emph{PCM and winding number---}By tuning the parameter $\Delta$, the merging process involves a phase transition from a semimetal to an insulator, where two massless Dirac points of opposite chirality combine into a topologically trivial gapped point. In its most general form, a gapless Dirac Hamiltonian is of the form  
\begin{equation}
\label{eq16}
\mathcal H_D(\vec{q})=\vec{q}\cdot( \vec{v}_1\sigma_x+\vec{v}_2\sigma_y).
\end{equation}
The Berry connection is defined as $ A_i=\langle u_\mathbf p |  \partial_i |u_\mathbf p\rangle$, where $|u\rangle$ is the Hamiltonian's eigenstate. For a Dirac point, the winding number is $w=\oint \mathbf A\cdot d\mathbf l$, in which the integration loop circles around the degenerate point. For the Hamiltonian Eq.~\eqref{eq16}, the winding is derived as $w=\lambda\text{sgn}(\vec{v}_1\wedge\vec{v}_2)$, where $\lambda=\pm 1$ is the band index and the wedge product here is defined as $\vec v_1 \wedge \vec v_2\equiv (\vec v_1\times \vec v_2 )_z$. Therefore, we have the relationship between the  winding number and the velocities ($\lambda=1$)
\begin{equation}
\begin{split}
w=\text{sgn}(\vec{v}_1\wedge\vec{v}_2)= \left \{
\begin{array}{ll}
    +1,                    & \text{for} \quad v_{1x}v_{2y} > v_{1y}v_{2x}\\
    -1,                     & \text{for} \quad v_{1x}v_{2y} < v_{1y}v_{2x}.
\end{array}
\right.
\end{split}
\end{equation}
Similarly, we consider the evolution of the wave packet in the form Eq.~\eqref{eq6}  with $\phi=(c_1,c_2)^T$ and $k_1=k_2=0$. In the large width limit, the spatial dynamics is totally determined by the Hamiltonian $\mathcal H_D(\mathbf q=0)$. The PCM  accordingly can be obtained as
\begin{equation}\label{xybart}
\begin{split}
\bar{x}(t)&=\left[(c_1c_2^*+c_1^*c_2)v_{1x}+i(c_1c_2^*-c_1^*c_2)v_{2x}\right]t,\\
\bar{y}(t)&=\left[(c_1c_2^*+c_1^*c_2)v_{1y}+i(c_1c_2^*-c_1^*c_2)v_{2y}\right]t.
\end{split}
\end{equation}
One can then find  two  initial spinor states, for example, $|\phi_1 \rangle=\frac{1}{\sqrt{2}}(1,1)^T$ and $|\phi_2 \rangle=\frac{1}{\sqrt{2}}(1,i)^T$, such that the corresponding PCM of wave packets take a simple form
\begin{equation}
\bar{x}_1=v_{1x}t, \quad
\bar{y}_1=v_{1y}t,
\end{equation}
and
\begin{equation}
\bar{x}_2=v_{2x}t,  \quad
\bar{y}_2=v_{2y}t,
\end{equation}
respectively. In a straightforward manner,  we rewrite the winding number in terms of PCM  as
\begin{equation}
\begin{split}
w=\text{sgn}(\bar{x}_1\bar{y}_2-\bar{x}_2\bar{y}_1)= \left \{
\begin{array}{ll}
    +1,                    & \text{for} \quad  \bar{x}_1\bar{y}_2>\bar{x}_2\bar{y}_1\\
    -1,                     & \text{for}  \quad \bar{x}_1\bar{y}_2<\bar{x}_2\bar{y}_1
\end{array}.
\right.
\end{split}
\end{equation}
Numerically, we apply the above method to demonstrate the winding number of the four Dirac points when $\Delta=0$,  in which case four different Dirac points located at $(0,0)$, $(2m^*c,0)$, $(-m^*c,-\sqrt{3}m^*c)$ and $(-m^*c,\sqrt{3}m^*c)$. Two of them are plotted in Fig.~\ref{fig6}, from which the winding number can be directly inferred. 

In the case that $\vec{v}_1=(v_{1x},0)$  $\vec{v}_2=(0,v_{2y})$, which is commonly seen in topological insulators. Via Eq.~\eqref{xybart}, the PCM now is 
\begin{equation}\label{xybart2}
\begin{split}
\bar x(t)=&2R\cos\theta v_{1x}t\\
\bar y(t)=&2R\sin\theta v_{2y}t
\end{split}
\end{equation}
where $c_1^*c_2 \equiv Re^{i\theta}$. The winding number around the Dirac point simply  is
\begin{equation}
\begin{split}
w=\text{sgn}(v_{1x}v_{2y})= \left \{
\begin{array}{ll}
    +1,                    & \text{for} \quad  v_{1x}v_{2y}>0\\
    -1,                     & \text{for}  \quad v_{1x}v_{2y}<0.
\end{array}
\right.
\end{split}
\end{equation}
Thus, we can determine the sign of $(v_{1x},v_{2y})$ and obtain the winding number by analyzing the PCM's motion direction with respect to the initial state parameter $\theta$.

\emph{Spin texture and winding number---}Within the topological band theory, the winding number can also be viewed as a mapping between manifolds. Rewriting the Hamiltonian Eq.~\eqref{eq16} in the form of  $\mathcal H_D=h\vec n(\mathbf k)\cdot \sigma_i$, the associated winding number  is~\cite{Duca2015} 
\begin{equation}
w=\frac{1}{\pi}\oint_\mathcal C n_x d  n_y, 
\end{equation}
where $\oint_\mathcal C$ denotes the integration along the loop circling around the Dirac point. The Bloch vector lies along the equator and the winding number $w$ characterizes the mapping from the momentum loop to the wave function space $\vec n$. The Hamiltonian can be seen as a magnetic field acting on a spin while its direction $\vec n$  varies in the momentum space. The number of times that the vector $\vec n$ winds along the equator gives the winding number $w$. 

\begin{figure}[t] \centering
\includegraphics[height=4.2cm]{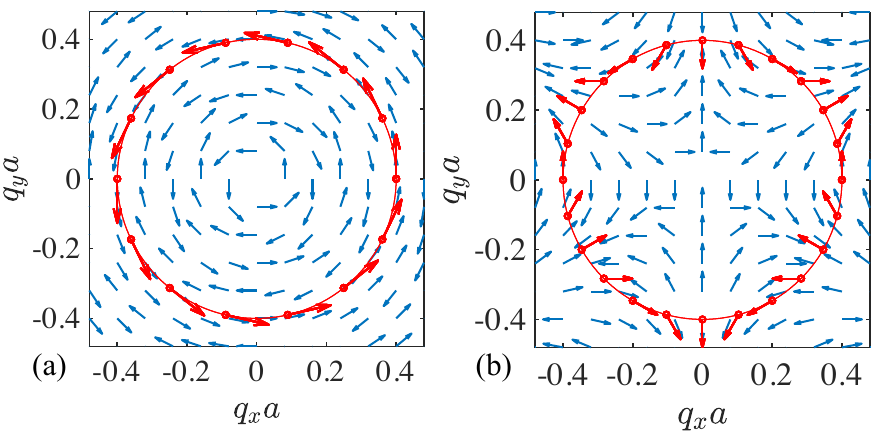}
\caption{ The spin texture for (a) the Dirac point and (b) the parabolic point with winding number $w=1$ and $w=2$, respectively. The width of wave packet is $d=5a$.  }
\label{spintexture}
\end{figure}

Here the initial state wave function is a Gaussian wave packet with a finite width and a spin-up spinor throughout momentum space. The evolution of the wave packet is the collection of spin's precession around $\vec n$ at each momentum. The initial spinor state is perpendicular to the $\vec n$, hence the spin will always be perpendicular to it during the evolution.  By examining the spin texture of the wave packet on isoenergy surfaces, one can derive the winding of $\vec n$ from the spin texture information. 
For example, considering the case of $\Delta=0$ and $c=0$ in Eq.~\eqref{eq3}, the corresponding Hamiltonian reads 
\begin{equation}
\mathcal H_V=
\begin{bmatrix}
0 &-\frac{q^2}{2m^*}e^{i2\varphi}\\
-\frac{q^2}{2m^*}e^{-i2\varphi} & 0
\end{bmatrix},
\end{equation}
where $q^2=q_x^2+q_y^2$ and $\tan \varphi=q_y/q_x$. The spin texture, $\langle \phi_q | \vec \sigma(t)| \phi_q \rangle$ ,  projected in the $xy$-plane can be obtained as
\begin{equation}
\begin{split}
s_x(\mathbf q,t)&=\sin(2\varphi)\sin(2Et),\\
s_y(\mathbf q,t)&=\cos(2\varphi)\sin(2Et).\\
\end{split}
\end{equation}
In Fig.~\ref{spintexture}, we plot the spin texture of both the Dirac point and the parabolic point. As only the direction of the spin texture determines the winding number, the overall time-dependent factor has been ignored. 

\emph{Experimental Implementation---}Finally, we discuss the experimental implementation of the initial state preparation and detection. In the cold atom system, the Bose gas confined in a harmonic trap can be approximately modeled by a Gaussian function, which naturally serves as a wave packet for simulation purposes here. By applying a magnetic field gradient~\cite{Tarruell2012} or moving the optical lattice~\cite{Fallani2004,Kling2010}, the atomic cloud can be accelerated such that the quasimomentum increases linearly up to the expected value.  After releasing the trap and switching on the optical lattice potential, the atomic ensemble would start expanding in the 2D honeycomb lattice. The time-evolving density profile can then be measured by the absorption imaging technique, hence allowing for the measurement of PCM~\cite{njp15073011}. For the measurement of the spin texture, the distribution  $\vec s(\mathbf q)$ can be experimentally determined from the time-of-flight images after the optical lattice is switched off, through which the momentum density distribution can be obtained.  By the Raman impulse, the $s_x(\mathbf q)$ and $s_y(\mathbf q)$ can be mapped to $s_z(\mathbf q)$ which is directly related to the (pseudo)spin-up and (pseudo)spin-down density $s_z(\mathbf q)=\frac 1 2[n_{\uparrow}(\mathbf q)+n_{\downarrow}(\mathbf q)]/[n_{\uparrow}(\mathbf q)+n_{\downarrow}(\mathbf q)]$~\cite{Alba2011,Wu2016,Hasan2022}.

The preparation of the initial spinor state can be easily achieved by utilizing the light-atom interaction if the atomic internal state is associated with the spin degree of freedom. In the graphene model here, the spin refers to the sublattice. When the width of the wave packet is much larger than the lattice constant,  the neighboring sites are equally populated and thus the initial spinor takes the form of $(1,1)$~\cite{Shen2019}. For any other spinor state, it can be prepared by switching on the optical lattice and driving the atoms to a proper quasimomentum. By using the lattice Hamiltonian near the quasimomentum, the desired spinor can then be prepared for a specific holding time. Since now the wave packet engages with only the spatial degrees of freedom,  the spin texture of the wave packet can then be resolved through the band tomography method~\cite{Flaeschner2016,Li2016a}.

\emph{ Summary---}In summary, we have investigated the wave packet dynamics of the tight-binding graphene model, which includes a third-nearest-neighboring coupling. The motion of Dirac points leads to band structures with gapped or gapless hybrid points. Corresponding wave packet dynamics show directed ZB motion and $C_3$ symmetry, indicating the linear-quadratic and $C_3$ symmetric dispersion, respectively. Besides, we have built up a relationship between the winding number of the Dirac point and the motion of the wave packet, and we also show that the winding numbers of both the Dirac point and parabolic points can be characterized by the spin texture of the wave packet. 

The wave packet dynamics elucidated has been extensively performed for the simulation of relativistic phenomena in various quantum and classical setups. Therefore, we could expect that our results benefit the experimental simulation of topological materials. Moreover, new topological states of matter such as the higher-order topological insulators and semimetals~\cite{Xie2021},  topological Euler insulators~\cite{Po2018, Peri2020}, and non-Hermitian topological insulators in open systems have recently attracted increasing interest~\cite{Okuma_2023}. Due to the similarity or novel features of the band structure, our results also provide a dynamical perspective for the exploration of the Dirac points~\cite{Herrera2023} or topological characteristics in those topological quantum states~\cite{quench3,quench2}.

\section*{Acknowledgements}
\label{Acknowledgements}
D.-D.L. and Z.L. acknowledge the support by the National Key Research and Development Program of China (Grant No. 2022YFA1405300), the National Natural Science Foundation of China (Grant No. 12074180), and the Guangdong Basic and Applied Basic Research Foundation (Grants No.2021A1515012350). X.S. acknowledges the support by the National Natural Science Foundation of China (Grant No. 12104430).


\bibliographystyle{apsrev4-1}
\bibliography{refs.bib}

\end{document}